\documentclass[DIV=calc, paper=a4, fontsize=11pt, twocolumn]{scrartcl}	 
\usepackage[english]{babel} 
\usepackage[protrusion=true,expansion=true]{microtype} 
\usepackage{amsmath,amsfonts,amsthm,amssymb,amsbsy} 
\usepackage[svgnames]{xcolor} 
\usepackage[hang, small,labelfont=bf,up,textfont=it,up]{caption} 
\usepackage{booktabs} 
\usepackage{fix-cm}	 
\usepackage{graphicx}	
\usepackage{sectsty} 
\usepackage{ulem}
\usepackage{soul} 
\allsectionsfont{\usefont{OT1}{phv}{b}{n}} 
\usepackage{cite}
\usepackage{fancyhdr} 
\usepackage{lastpage} 
\usepackage[breaklinks=true]{hyperref}
\usepackage{doi}

\hypersetup{
  colorlinks   = true, 
  urlcolor     = blue, 
  linkcolor    = blue, 
  citecolor   = red 
}
\usepackage{url}

\usepackage{lettrine} 
\newcommand{\initial}[1]{ 
\lettrine[lines=3,lhang=0.3,nindent=0em]{
\color{DarkGoldenrod}
{\textsf{#1}}}{}}

\renewcommand\emph[1]{{\it{#1}}}

\usepackage{titling} 

\newcommand{\HorRule}{\color{DarkGoldenrod} \rule{\linewidth}{1pt}} 

\pretitle{\vspace{-80pt} \begin{flushleft} \HorRule \fontsize{24}{24} \usefont{OT1}{phv}{b}{n} \color{DarkRed} \selectfont} 

\title{Weak-value amplification: state of play} 

\posttitle{\par\end{flushleft}\vskip 0.5em\vspace{-20pt}} 

\preauthor{\begin{flushleft}\large \lineskip 0.5em \usefont{OT1}{phv}{b}{sl} \color{DarkRed}} 

\author{George C. Knee $^{1*}$, Joshua Combes $^{2,3}$, Christopher Ferrie $^{2}$, and Erik M. Gauger $^{1,4}$} 
\postauthor{\footnotesize \usefont{OT1}{phv}{m}{sl} \color{Black} 
\, $^1$Department of Materials, University of Oxford, Oxford OX1 3PH, United Kingdom, $^2$Center for Quantum Information and Control, University of New Mexico, Albuquerque, New Mexico, 87131-0001, $^3$ ARC Centre for Engineered Quantum Systems, School of Mathematics and Physics, The University of Queensland, St Lucia, QLD 4072, Australia,  $^4$ SUPA, Institute of Photonics and Quantum Sciences, Heriot Watt University, Edinburgh EH14 4AS, United Kingdom, $^*$\href{mailto:gk@physics.org}{gk@physics.org}

\vspace{-20pt}
\par\end{flushleft}\HorRule} 

\date{\today} 

\begin{document}
 
\maketitle 

\thispagestyle{fancy} 


\initial{W}\textbf{eak values arise in quantum theory when the result of a weak measurement is conditioned on a subsequent strong measurement. The majority of the trials are discarded, leaving only very few successful events. Intriguingly those can display a substantial signal amplification. This raises the question of whether weak values carry potential to improve the performance of quantum sensors, and indeed a number of impressive experimental results suggested this may be the case. By contrast, recent theoretical studies have found the opposite: using weak-values to obtain an amplification generally worsens metrological performance. This survey summarises the implications of those studies, which call for a reappraisal of weak values' utility and for further work to reconcile theory and experiment.}


\section*{Weak measurements vs. weak values}

A quantum weak measurement is a procedure whereby only a little bit of information about a quantum system is obtained; as a consequence, the system is only disturbed a little. This is in contrast to the usual strong measurements, which give a lot of information but inflict a large disturbance on the system. 

Imagine the needle on a poor-quality analogue voltmeter, which twitches or deflects in response to an electrical signal. In a weak measurement, the amount of deflection is only loosely correlated with the true voltage---because the needle also twitches about randomly.  The expected value of the deflection, however, is precisely the true voltage: over many trials the average deflection will reveal the true voltage with increasing precision. 

Physically, a weak measurement can be implemented by introducing a weak interaction between the system of interest and an ancillary meter degree of freedom \cite{BarLanPro1982,CavMil1987}, a procedure known as a von Neumann measurement. Consider the following interaction Hamiltonian between system and meter:
\begin{align}
H = g A \otimes P~,
\end{align}
where $g$ is a scalar quantity, $A$ is the operator associated with the relevant system observable, and $P$ is an operator effecting a shift of the meter variable. 
The dynamics induced by this Hamiltonian (we assume it is switched on and off again instantaneously) will build up correlations between the system and meter. The degree of correlations, or the `measurement strength' can be controlled for example by changing $g$; with strong and weak limits being attained when $g \to \infty$ and $g \to 0$, respectively.
Owing to these correlations, a subsequent measurement on the meter alone may reveal information about the system. The back-action imparted to the system can be understood by examining the effective measurement operators of the system (sometimes called POVM, or Positive Operator Values Measure elements \cite{NielsenChuang2000}): strong measurement operators are projective, giving maximal information but also imparting the largest back-action. Intermediate strength measurements impart less back-action but give less information. In the limit of a vanishing interaction, the POVM elements become the identity matrix, and no measurement is performed at all. As long as a measurement of some finite strength is performed, however, the average deflection is precisely the true voltage, which will be revealed over many trials with increasing precision.

Weak measurements have become part of the standard toolbox in the modern field of quantum control: they can be a useful method of stabilising a quantum computation, where information must be prevented from leaking into the environment~\cite{GillettDaltonLanyon2010}. If a weak measurement is repeated on a single system enough times to provide the same information as a strong measurement, a comparable back action will be imparted. 

\begin{figure}[th!]
\includegraphics[width=\linewidth]{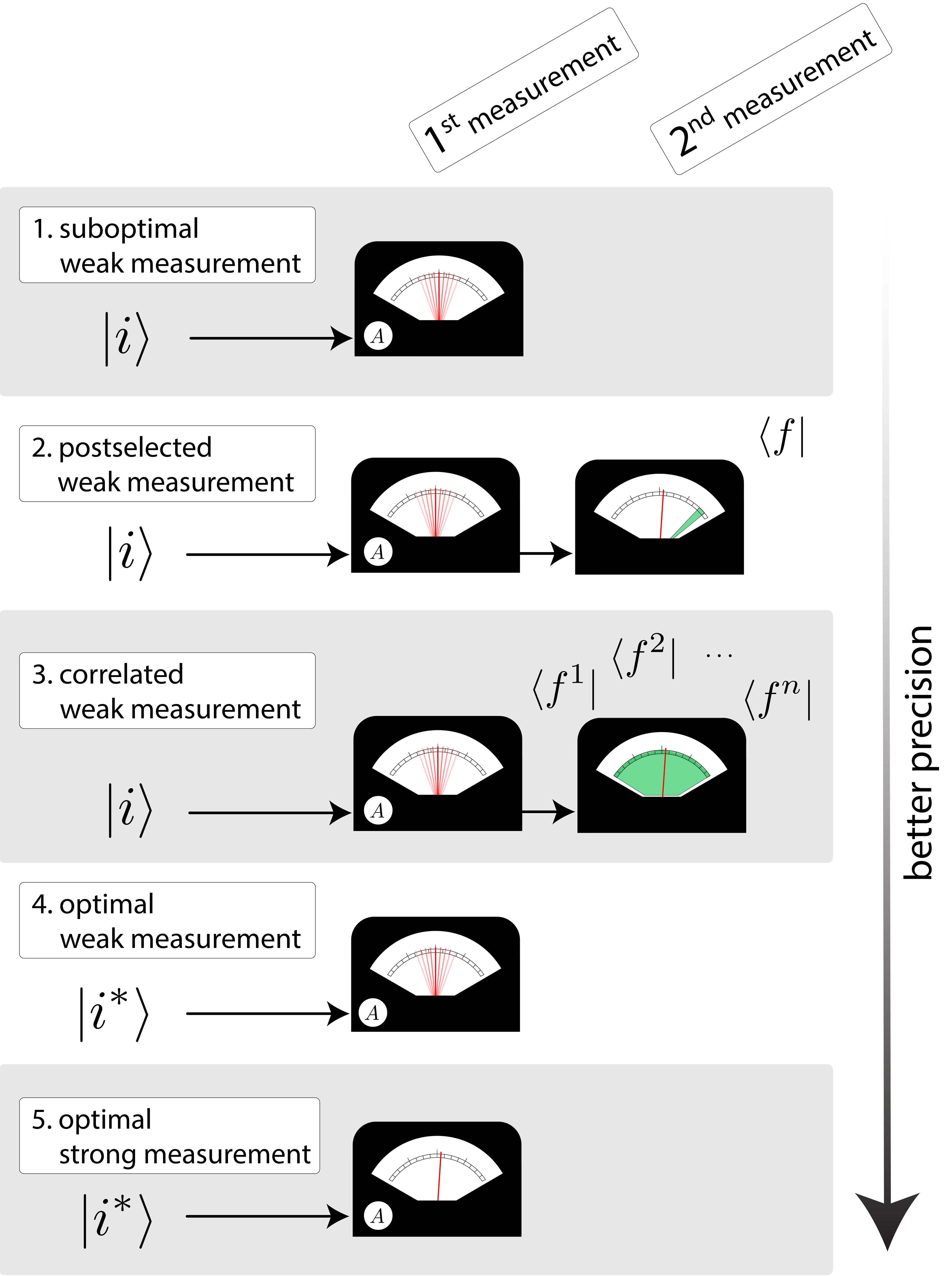}
\caption{\label{voltmeters}Various schemes for parameter estimation. The experimenter can choose the initial state $|i\rangle$, the amount of uncertainty in the measurement of observable $A$ (depicted here by a fluctuating meter needle), and also whether to correlate the first measurement with a second one (which projects onto a final states $|f^i\rangle$). The green areas on the second meter depict a value that must occur for the experiment to be successful -- otherwise it is rejected. It has recently been shown that a postselected weak measurement (2.) will not give more information than three simple alternatives, despite its association with large meter readings known as `anomalous weak values'. It is better to either (3.) keep all of the data from both measurements, or (4.) prepare an optimal initial state $|i^*\rangle$ and dispense with the second measurement altogether. A single strong measurement (5.) gives the most information of all. Scheme (1.) involves a suboptimal initial state and no conditioning, giving the worst performance of all.}
\end{figure}

The weak value is different to a weak measurement, but often defined in conjunction.  A weak value, like the expected value, is a well-defined quantity that arises from applying standard quantum mechanics to a particular measurement protocol involving pre and postselection~\cite{AharonovAlbertVaidman1988,DuckStevensonSudarshan1989}. The procedure to obtain a weak value is explained in detail in Ref.~\cite{DresselMalikMiatto2014}, but we sketch the idea here (see Figure~\ref{voltmeters}, case 2). First, the system of interest is prepared in a known initial state (e.g., a predetermined voltage) $|i\rangle$. It is measured weakly (for example, by using a poor quality voltmeter coupled to observable $A$) and then measured again, this time strongly (with a good quality meter). The first and second measurement are generally described by complementary observables. Finally, the data from the poor voltmeter are culled, a step known as postselection; only those instances where the second measurement reported a particular unlikely result are kept (for example the result corresponding to final state $|f\rangle$). The real part of the weak value
\begin{align}
A_w = \frac{\langle f | A| i \rangle}{\langle f | i \rangle}
\end{align}
is then (approximately) proportional to the expected value of the surviving data from the first measurement. When the reading on the second meter is a very rare one, the weak value can become quite large, and in particular larger than the largest eigenvalue  $\lambda^* = \max{|A|}$ of the observable. Intuitively, the anomalously large deflection seems to indicate the potential to increase or amplify the precision of measuring devices---why live with a small signal when a large one can be arranged? Intuition, however, is notoriously unreliable. 

\section*{The cost of amplification}
The weak-value approximately determines the magnitude of needle deflection, and it can be larger than the expected value
\begin{align}
\langle A \rangle = \langle i | A | i\rangle \leq \lambda^*~,
\end{align}
which provides an upper bound on the deflection when no second measurement is performed, whilst in the postselected case $A_w > \lambda^*$ is possible. This has motivated many experiments to convert tiny effects into larger ones using postselection, usually with the aim of estimating the coupling between two quantum degrees of freedom. The first such study was Onur Hosten and Paul Kwiat's 2008 experiment~\cite{HostenKwiat2008}, which detected the spin-Hall effect of light (a coupling between the polarisation and transverse momentum of light at an interface between media with different refractive indices) with postselected weak measurements. As with many experiments with weak values, the polarisation of a beam of light played the role of the quantum system and the deflection of the beam replaced the twitching of the meter needle -- other notable experiments include the measurement of a femtoradian mirror tilt~\cite{DixonStarlingJordan2009} and other small optical effects. However, the technique applies generally to different degrees of freedom in optical as well as in a range of other physical systems. 

The recent experimental successes being reported lend weight to the longstanding question whether weak value amplification can deliver a fundamental advantage for parameter estimation, or whether it should merely be regarded as a convenient experimental tool in certain circumstances.  In the last year, a number of researchers working independently have addressed the question of whether weak values do indeed unlock superior performance employing the rigorous framework of parameter estimation. Before describing the formalism of parameter estimation in order to understand those results, let us discuss the meaning of postselection and give two intuitive reasons (which can be made mathematically rigorous) why the amplification provided by weak values may not be a `silver bullet' for precision measurements. 

The first reason is that the data are necessarily extremely noisy; for the measurement to truly qualify as weak, the needle on the measuring device is continually wandering under quantum fluctuations. Any systematic deflection of the needle is, by design, hidden by the fundamental quantum uncertainty in any given run. Detecting the signal necessitates the use of a statistical approach. In any weak measurement, postselected or otherwise, a very large number of trials is vital for a significant conclusion to be reached. Only a strong measurement can provide a precise estimate after a single trial, when all sources of noise are eradicated.

The second reason is that the anomalously large deflections are very rare. The larger the `amplification' that is desired, the more trials are required before an experiment succeeds. In optical experiments, a low success probability translates into a much reduced photon detection rate. Therefore, the effect leads to an~\emph{attenuation} as much as it leads to an amplification. This latter point is known as the problem of low postselection success probability.

It is important to distinguish two notions of `postselection': On the one hand, it can be understood as a physical step -- for example the inclusion of a polarising filter which only allows certain photons to reach the detector. The data from the first measurement are then only recorded \emph{after} the second measurement has triggered the amplification. On the other hand, postselection may be understood as the rejection of certain events from a larger dataset -- the amplification is then an artefact of data processing only. This distinction matters as the two cases lead to different ways of benchmarking the weak value technique,  which can in turn give rise to slightly different conclusions.

Under the first definition, the weak value amplification approach and its standard benchmark strategy correspond to (slightly) different physical setups and protocols: one compares the use of postselection (case 2) against not using a second measurement at all (case 4 in Figure~\ref{voltmeters}). This is the approach taken by Refs.~\cite{KneeGauger2014,Kedem2012,JordanMartinez-RinconHowell2014,PangBrun2014}. 

Taking the second definition, the experimental setup remains identical (i.e.~a second strong measurement is always performed) and postselection is understood as partitioning the recorded data into two sets, only one of which will be used for the parameter estimation process at hand. Then, one effectively compares case 2 with case 3 in Figure~\ref{voltmeters}. This notion of postselection is employed by Refs.~\cite{TanakaYamamoto2013,FerrieCombes2014,CombesFerrieJiang2013,ZhangDattaWalmsley2013}.

\section*{Parameter estimation}
\begin{figure}[b]
\includegraphics[width=0.96\linewidth]{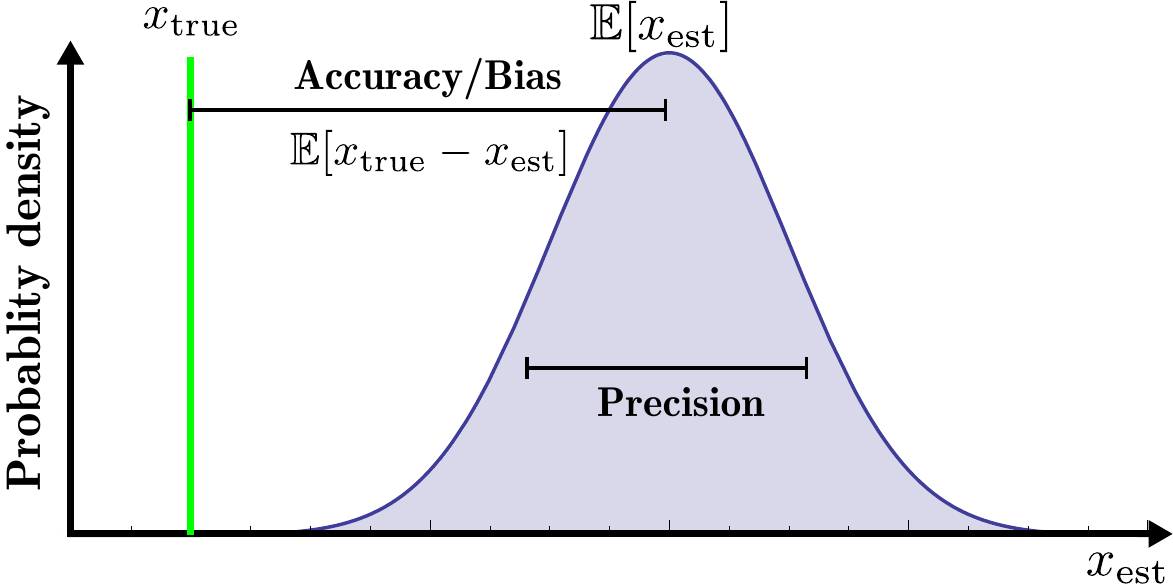}
\caption{\label{fig_est}An unbiased estimator has the property that in an infinite number of trials its average value will equal the true value. The quality of an unbiased estimator is characterized by the variance of the estimator. The smaller the variance the more precise the estimator is.}
\end{figure}

Gathering and interpreting data is the very essence of empirical science. For results to be meaningful, they should correspond accurately to those predicted by a theoretical model, and a statement about their uncertainty must be made. When writing up the results of an experiment, one states the {\em estimate} of the measured quantity along with an {\em uncertainty} -- the voltage was $5.0$V $\pm0.1$V, for example. Both numbers are the output of statistical calculations. Consider estimating the voltage of a constant signal using a noisy voltmeter. The true voltage is denoted by $x_{\rm true}$. Typically one looks at the voltmeter and records the (random) value $x_i$, the needle deflection in the $i$'th experiment.  An~\emph{estimator} is a function of the data that produces an estimate for the true voltage. Often, the average of $x$ over $N$ trials is a good choice, $x_{\rm est}=(1/N)\sum_i^N x_i$.

If the experimenter happens to be sitting at an angle to the voltmeter dial, she might consistently over- or underestimate the deflection. This will result in a biased estimate of the voltage. An unbiased estimator, on the other hand, is perfectly accurate on average $\mathbf{E}[ x_{\rm true} - x_{\rm est}]=0$, where the symbol $\mathbf{E}$ denotes the expected value. Aside from accuracy, the quality of the estimation procedure is characterized by the precision, defined as the variance of the estimator. See Figure~\ref{fig_est} for a simple illustration.

In many instances it can be difficult to directly calculate the precision theoretically. A mathematical tool known as the Fisher information \cite{vanTrees1968,Wasserman2004} allows one to place an lower bound on the variance of an estimator. The Fisher information gives a single number $F$, providing a powerful link between theory and experiment. If one processes the data from the experiment with the best possible, or optimal estimator, the variance will be given by the inverse of the Fisher information. That means the experimenter will report e.g. $5.0$V $\pm (1/\sqrt{NF})$V. Clearly a higher $F$ is better because it implies a lower uncertainty in the estimate. The magnitude of the Fisher information depends on exactly how the experiment is performed. It thus provides an excellent way of comparing different approaches to parameter estimation. The dependence on $N$ means that, as long as the experiment is repeatable, the uncertainty can become arbitrarily small by increasing the number of trials.

\section*{An estimation inequality}

Applying the above framework of parameter estimation to the case of weak value amplification, several recent theoretical studies have reached the conclusion that, given the same number of input resources, a weak-value strategy will generally not outperform the standard metrology strategy~\cite{KneeBriggsBenjamin2013,TanakaYamamoto2013,KneeGauger2014,FerrieCombes2014,CombesFerrieJiang2013,ZhangDattaWalmsley2013}. These results are essentially all captured in an inequality constraining the corrected Fisher information:
\begin{align}\label{fish_ineq}
  p(\checkmark) F_{\rm weak\ value\ } \le F_{\rm standard}~,
\end{align}
where $p(\checkmark)$ is the postselection success probability. As discussed above, the standard Fisher information could be one of a number of alternatives (see Figure~\ref{voltmeters}). For example keeping all of the data (rather than discarding most of it) will give higher Fisher information~\cite{TanakaYamamoto2013,FerrieCombes2014,StrubiBruder2013}; similarly, choosing an optimal initial state and not performing the second measurement at all will outperform a postselected weak measurement~\cite{KneeGauger2014}. A strong measurement, if available, provides the highest precision of all~\cite{TanakaYamamoto2013,FerrieCombes2014,CombesFerrieJiang2013,ZhangDattaWalmsley2013,KneeGauger2014}. Furthermore, a simple estimator based upon the approximate weak value (rather than the true average of the postselected weak measurement) is not unbiased: there is a systematic error in the estimate in any real experiment. Accuracy and precision are thus both worse when postselection is used. 

It is possible, however, for $p(\checkmark)F_{\rm weak\ value}$ to approach equality with $F_{\rm standard}$ in a restricted parameter regime. With weak-values, a simple estimator can match the accuracy of the standard strategy only when the measurement is infinitely weak, and the precision can match the standard strategy only when the postselection probability is zero. Neither of these is physically possible, although both can be approximated with arbitrary closeness: the fact that only a negligible amount of information is discarded through postselection has been confirmed in the laboratory~\cite{VizaMartinez-RinconAlves2014}. It is therefore interesting to consider how the two approaches compare away from the ideal scenario. Below we list some of the ongoing investigations.

\section*{Estimation in non-ideal scenarios}
The true limits to parameter estimation in the laboratory stem from a variety of sources other than the inherent quantum uncertainty. Are there scenarios where the above results do not apply, and where an advantage can be found?

\subsection*{Getting lucky} 

Sometimes $F_{\rm weak\ value\ }\ge F_{\rm standard}$ and postselection might temporarily provide more information, because the inequality in (\ref{fish_ineq}) only applies when the number of trials is large. However, the probability of a windfall persisting decreases exponentially as more trials are performed~\cite{FerrieCombes2014,CombesFerrieJiang2013}.

\subsection*{Technical noise} 

Experiments exploiting different physics can exhibit varying robustness to noise and detector imperfections. This fact has motivated a long standing feeling that weak-value amplification can help to overcome such problems. However, we have shown that estimation inequalities analogous to~(\ref{fish_ineq}) hold in the presence of a broad class of noise before, during, and after the weak measurement~\cite{CombesFerrieJiang2013,FerrieCombes2014,KneeGauger2014,KneeBriggsBenjamin2013}. Together these articles treat the most prevalent types of noise (including pixelation, detector jitter, and dephasing), and a similar approach can be used to analyse other imperfections. For instance, Jordan {\it et al.}~have claimed that for specific noise models a weak-value-amplified technique gives higher Fisher information than conventional methods~\cite{JordanMartinez-RinconHowell2014}. By engineering these noise models artificially, these predictions have been explored experimentally~\cite{VizaMartinez-RinconAlves2014}.

\subsection*{Suboptimal estimation strategies} 

Some researchers question the use of optimal estimators, and therefore question the relevance of the Fisher information to actual experiments. A recent manuscript imagines a type of noise that cannot be modelled statistically, and hence affects the uncertainty in the estimate in a more profound way~\cite{LeeTsutsui2014}. Refs.~\cite{JordanMartinez-RinconHowell2014, JordanTollaksenTroupeDresselAharonov2014} argue that time-correlated noise, for example, renders the optimal estimator impractical due to computational complexity. This is not, by itself, a compelling argument for employing weak values, however: one must show that the entire weak value protocol, including the post processing, is not outperformed by a suitable benchmark strategy. Further, it has been shown that a computationally trivial estimator exists for time correlated noise and outperforms the weak-value-amplification technique \cite{FerrieCombes2014}. If one has chosen a sub-optimal initial state $|i\rangle\neq|i^*\rangle$, then performing an optimal postselection $\langle f^*|$ can lead to more (corrected) Fisher information than not performing the second measurement at all (case 1)~\cite{PangBrun2014} -- otherwise, when $|i^*\rangle$ is chosen, the Fisher information of case 4 achieves the theoretical maximum and cannot be surpassed by any pre and postselected strategy, including an optimal one.

\subsection*{Imaginary weak values} 

Imaginary weak values occur when the postselected weak measurement is performed in a certain way: the amplification is then seen in Fourier space, rather than real space. 

Whether or not imaginary weak values merit special consideration depends on what is meant by `postselection'. Comparing cases 2 and 4 in Figure~\ref{voltmeters}, i.e. protocols with different \emph{physics}, there may be  scope for altering the apparatus in a fashion which tailors it to better fit the available hardware. In a time-domain experiment, for example, a frequency analyser is sometimes preferred to a stopwatch~\cite{Kedem2012,JordanMartinez-RinconHowell2014}. However, unless there is a severe mismatch between the quality of detection in the two variables, imaginary weak values will not provide a significant advantage~\cite{KneeGauger2014}.
On the other hand, when case 2 is compared to case 3 in Figure~\ref{voltmeters}, then postselection is merely data rejection, which cannot improve estimation under the most general evolution allowed by quantum theory~\cite{CombesFerrieJiang2013}. This latter analysis therefore covers experiments involving any defined quantities, including imaginary and even complex weak values. 

\subsection*{The true cost of estimation} 

It is interesting to consider the different notions of the cost associated with an investigation. One must spend time and energy to perform the experiment; there is a financial cost accompanying the hardware; and a computational cost associated with data processing and estimation.

Weak value estimators have been conjectured to offer a computationally simpler alternative than standard techniques, despite the extra apparatus required~\cite{JordanMartinez-RinconHowell2014}. 
Although this is an appealing idea, standard methods are often equally cheap to compute as the weak-value estimator whilst being more precise~\cite{FerrieCombes2014}. For example, for case 4 in Figure~\ref{voltmeters} the optimal estimator is simply proportional to the average measurement result. 
One should bear in mind that performing unbiased estimation with weak values requires more complicated postprocessing: one must account for non-linear effects not captured in the weak value, which is only an approximate quantity.

Another type of cost arises uniquely in optics experiments. Lasers can easily emit $10^{10}$ photons per nanosecond, making the creation-cost per photon almost negligible. The detection-cost of a photon, by contrast, is often effectively much higher, especially if the photodetector saturates very quickly~\cite{Vaidman2014}. A variant accounting philosophy, which weighs output resources more heavily than input resources, may give rise to a different conclusion to the one reached above~\cite{DresselBrunKorotkov2014,CombesFerriePreparation}. Interestingly, it is in exactly these special circumstances (those of large numbers of cheap input photons) that the weak-value phenomenon is known to have a classical explanation. By contrast, genuinely quantum-enhanced metrology typically exploits effects involving~\emph{single quanta}~\cite{GiovannettLloydMaccone2004}, and then the number of input resources becomes the limiting factor.

Experimental runs which fail the postselection test can be  ``recycled'' for a further interaction with the unknown parameter~\cite{DresselLyonsJordan2013}, and this procedure repeated until they eventually pass. If the runs (typically thought of as photons) are reused in this way but are not~\emph{recounted} as a resource, the technique is superior to a one-interaction-per-photon method. This is of no great surprise, because the interaction with the unknown parameter is precisely when the information is imparted to the probe, and thus increasing the total number of interactions is worthwhile. Recycling is therefore a technique to increase Fisher information given a fixed laser input power, for example. It is of course possible to recycle in the `standard' strategies, by rerouting all photons (not just those that failed postselection).

Finally it is worth noting that experiments are frequently performed far from the ultimate performance limit. Some decisions are made in the laboratory merely for convenience, for example to cope with incidental factors. From this point of view, weak-value amplification may be judged `handy' on a case-by-case basis.

\subsection*{Combination with other quantum effects} 

Another direction that has begun to be investigated is combining weak value approaches with more established quantum metrology methods (see e.g. \cite{Paris09,TothApel14,DemJarKol14}); for example with entanglement \cite{PangDresselBrun2014} or squeezing~\cite{PangBrun2014}. Both these effects are known to unlock a genuine quantum advantage for certain metrological tasks, but at this point it is not clear whether combining them with weak values can indeed lead to anything greater than the sum of its parts. 

\section*{Conclusion}

For a number of years, experimental successes involving the weak-value technique have lent credence to the intuition that an `amplified' signal is always a good thing. By contrast, recent theoretical results cast doubt on whether weak value techniques can really offer a fundamental quantum advantage for metrology. However, the debate continues since the conclusion one reaches depends on various subtleties such as the the resource accounting philosophy, the choice of benchmark, and the notion of postselection employed. In any case, however, we here submit that, by itself, a large amplification factor is not sufficient for an advantage in parameter estimation.

\section*{Acknowledgements}
We are grateful to Carl Caves, Animesh Datta, Yaron Kedem, and Jonathan Leach for their feedback on this document. GCK was supported by EPSRC Grant No. EP/P505666/1. JC and CF were supported in part by NSF Grant Nos. PHY-1212445 and PHY-1314763. JC was also supported by the Australian Research Council Centre of Excellence for Engineered Quantum Systems grant number CE110001013. CF was also supported in part by the Canadian Government through the NSERC PDF program. EMG acknowledges support from the John Templeton World Charity Foundation "Experimental Tests of Quantum Reality" project and the Royal Society of Edinburgh / Scottish Government. 

\bibliographystyle{plainurl}
\bibliographystyle{plain}

\end{document}